\documentclass[preprint,showpacs,preprintnumbers,amsmath,amssymb]{revtex4}

\def\undersim#1{\setbox9\hbox{${#1}$}{#1}\kern-\wd9\lower
    2.5pt \hbox{\lower\dp9\hbox to \wd9{\hss $_\sim$\hss}}}
\usepackage{amssymb}
\usepackage{amsmath}
\usepackage{graphicx}
\usepackage{color}

\def\undersim#1{\setbox9\hbox{${#1}$}{#1}\kern-\wd9\lower
    2.5pt \hbox{\lower\dp9\hbox to \wd9{\hss $_\sim$\hss}}}

\def\mv{{\mathbf v}}
\def\mr{{\mathbf r}}

\def\mr{{\mathbf r}}

\def\mk{{\mathbf k}}

\begin{document}

\title{Squeezed spectra of bosons and antibosons with different
in-medium masses}

\author{Yong Zhang$^{1,}$\footnote{zhy913@jsut.edu.cn}}
\author{Hui-Qiang Ding$^{2}$}
\author{Shi-Yao Wang$^{2}$}
\affiliation{\footnotesize $^1$School of Mathematics and Physics, Jiangsu University of Technology, Changzhou, Jiangsu 213001, China\\
$^2$School of Physics, Dalian University of Technology, Dalian, Liaoning 116024, China}

\date{\today}

\begin{abstract}
We study the influence of the in-medium mass difference between boson and antiboson on their spectra.
The in-medium mass difference may lead to a difference between the transverse momentum spectra
of boson and antiboson. This effect increases with the increasing in-medium mass difference between boson and antiboson.
The difference between the transverse momentum spectra of boson and antiboson
increases with the increasing expanding velocity of the source and decreases
with the increasing transverse momentum in large transverse mass region ($m_T\,>\,1.6$ GeV).
{\color{black}The interactions between the hadron and the medium may increase with the increasing temperature of the medium and the
higher freeze-out temperature may lead to a larger mass difference between boson and antiboson, and may give rise to a larger difference
between the transverse momentum spectra of boson and antiboson for higher freeze-out temperature.}


\end{abstract}

\maketitle

\section{Introduction}
The transverse momentum spectra of particles are important observables in high-energy heavy-ion collisions.
They can be utilized to investigate the thermalization and expansion of the systems produced in such
collisions \cite{STAR_PRL04_TMS,PHENIX_PRC04_TMS,ALICE_PRL12_TMS,ALICE_PRC13_TMS}.

The in-medium mass shifts of bosons may cause a squeezing effect, and directly lead to a squeezed back-to-back
correlation between boson and antiboson \cite{AsaCso96,AsaCsoGyu99,Padula06,Zhang15a}. This squeezing effect is
related to the in-medium mass of the bosons, through a Bogoliubov transformation between the creation
(annihilation) operators of the quasiparticles in the medium and the corresponding free particles \cite{AsaCso96,AsaCsoGyu99,Padula06,Zhang15a}.
The squeezing effect also affects the transverse momentum spectra of bosons \cite{AsaCsoGyu99,Padula06}.
The influence of squeezing effect caused by in-medium mass shift on transverse momentum spectra and elliptic
flow of $\phi$ meson was studied in previous work \cite{Zhang20}. Since the antiparticle of $\phi$ meson is itself,
the in-medium masses of $\phi$ meson and its antiparticle are treated as the same \cite{Zhang20}. The interactions of charged particles and
their corresponding antiparticles in a medium are different \cite{{LiLeeBrown-NPA97,SibTsuTho-EPJA99,CMKo-JPG01,
Mishra-PRC69-04,Mishra-PRC70-04,BlaCosKal-PRD12,Mishra-EPJA2019}}. The in-medium mass difference between a boson
and an antiboson may lead to different effects on their spectra. It is necessary to study the squeezing effect on spectra of boson
and antiboson with different in-medium masses.

This paper is organized as follows. In Sec.II, we present the formulas of the single-particle
momentum distributions for boson and antiboson with different in-medium masses. Then, the squeezed spectra of
boson and antiboson are shown in Sec. III.
Finally, summary and discussion of this paper are given in Sec. IV.

\section{FORMULAS}
Denote $a_\mk\, (a^\dagger_\mk)$ as the annihilation (creation) operator of the free boson
with momentum $\mk$ and mass $m_a$, and $b_\mk\, (b^\dagger_\mk)$ as the annihilation (creation)
operator of the free antiboson with momentum $\mk$ and mass $m_b$. For a pair of free boson and antiboson,
$m_a\,=\,m_b=\,m$.
The invariant single-particle momentum distributions of boson and antiboson
can be expressed by
\begin{equation}
N_a(\mk)=\omega_{a,\mk}\frac{d^{3}N}{d\mk}=\omega_{a,\mk}\langle a^\dagger_{\mk} a_{\mk} \rangle,
\end{equation}
\begin{equation}
N_b(\mk)=\omega_{b,\mk}\frac{d^{3}N}{d\mk}=\omega_{b,\mk}\langle b^\dagger_{\mk} b_{\mk} \rangle,
\end{equation}
where $\langle \cdots \rangle$ means the thermal average, and $\omega_{a,\mk}=\sqrt{\mk^2 + m^2}$
and $\omega_{b,\mk}=\sqrt{\mk^2 + m^2}$ are the energy of the free boson and antiboson, respectively.

Denote $a'_\mk\, (a'^\dagger_\mk)$ as the annihilation (creation) operator of the boson with
momentum $\mk$ in medium, and $b'_\mk\, (b'^\dagger_\mk)$ as the annihilation (creation)
operator of the antiboson in medium.
Assuming the energy split between the boson and antiboson in the medium is $2\delta$, the operators $(a_\mk, a^\dag_\mk,
b_\mk, b^\dag_\mk)$ for the free particles and $(a'_\mk, a'^\dag_\mk, b'_\mk,
b'^\dag_\mk)$ for the quasiparticles were related by the Bogoliubov transformation \cite{XPZ-19}
\begin{equation}
a_\mk =c_\mk a'_\mk +s^*_{-\mk} b'^\dag_{-\mk}, ~~~~
b_\mk ={\bar c}_\mk b'_\mk +{\bar s}^*_{-\mk} a'^\dag_{-\mk},
\end{equation}
where
\begin{equation}
c^*_{\pm\mk}=c_{\pm\mk}={\bar c}^*_{\pm\mk}={\bar c}_{\pm\mk}=\cosh r_\mk,
\end{equation}
\begin{equation}
s^*_{\pm\mk}=s_{\pm\mk}={\bar s}^*_{\pm\mk}={\bar s}_{\pm\mk}=\sinh r_\mk,
\end{equation}
\begin{equation}
r_\mk =\frac{1}{2} \ln\left({\omega_\mk}/{\Omega_\mk}\right),
\end{equation}
\begin{equation}
\Omega_\mk =\sqrt{\mk^2 +m'^2 +\delta^2}.
\end{equation}
The in-medium masses of the boson and antiboson are:
\begin{equation}
m'_{a}= (\Omega_{\mk} + \delta)\big|_{\mk=0}=\sqrt{m'^2+\delta^2}
+ \delta,
\end{equation}
\begin{equation}
m'_{b}= (\Omega_{\mk} - \delta)\big|_{\mk=0}=\sqrt{m'^2+\delta^2}
- \delta,
\end{equation}
where $m'$ is the in-medium mass of the boson or antiboson for $\delta\,=\,0$.
For a pair of boson and antiboson, the subscript $a$ indicates the particle with larger mass and the subscript $b$
represents the particle with smaller mass.

For a homogeneous source with volume $V$ and temperature
$T$, the single particle momentum distributions of boson and antiboson can be expressed by \cite{AsaCsoGyu99,Padula06,XPZ-19}
\begin{equation}
N_a(\mk)=\frac{V}{{\color{black}(}2\pi{\color{black})}^3}\omega_\mk \Bigl\{|c_{\mk}|^2\,n_{a,\mk}+|s_{-\mk}|^2(n_{b,-\mk}+1)\Bigl\},
\end{equation}
\begin{equation}
N_b(\mk)=\frac{V}{{\color{black}(}2\pi{\color{black})}^3}\omega_\mk \Bigl\{|{\bar c}_{\mk}|^2\,n_{b,\mk}+|{\bar s}_{-\mk}|^2(n_{a,-\mk}+1)\Bigl\},
\end{equation}
\begin{equation}
n_{a,\mk}=\frac{1}{\exp[(\Omega_{\mk}+\delta)/T]-1},
\end{equation}
\begin{equation}
n_{b,\mk}=\frac{1}{\exp[(\Omega_{\mk}-\delta)/T]-1}.
\end{equation}

For hydrodynamic sources, the single particle momentum distributions of boson and antiboson
become \cite{AsaCsoGyu99,Padula06,Zhang15a,YZHANG_CPC15,Cooper,KolHei03,Kol00}
\begin{eqnarray}\label{SP}
 N_a(\mk)\!&&=\!\int \frac{g_i}{(2\pi)^3}d^4\sigma_{\mu}(r)k^\mu\, \! \Bigl\{|c'_{\mk'}|^2\,
n'_{a,\mk'}+\,|s'_{-\mk'}|^2\,[\,n'_{b,-\mk'}+1]\Bigr\},
\end{eqnarray}
\begin{eqnarray}\label{SP1}
 N_b(\mk)\!&&=\!\int \frac{g_i}{(2\pi)^3}d^4\sigma_{\mu}(r)k^\mu\, \! \Bigl\{|c'_{\mk'}|^2\,
n'_{b,\mk'}+\,|s'_{-\mk'}|^2\,[\,n'_{a,-\mk'}+1]\Bigr\},
\end{eqnarray}
\begin{equation}\label{csk}
c'_{\pm\mk'}=\cosh[\,f'_{\mk'}\,], \,\,\,s'_{\pm\mk'}=\sinh[\,f'_{\mk'}\,],
\end{equation}
\begin{eqnarray}
f'_{\mk'}=\frac{1}{2} \log \left[\omega'_{\mk'}/\Omega'_{\mk'}\right]
=\frac{1}{2}\log\left[k^{\mu}u_{\mu}(r)/k^{*\nu}u_{\nu}(r)
\right],
\end{eqnarray}
\begin{eqnarray}
&&\hspace*{-7mm}\Omega'_{\mk'}(r)=\sqrt{\mk'^2(r)+m'^2 +\delta^2}\nonumber\\
&&\hspace*{3.9mm}=\sqrt{[k^{\mu} u_{\mu}(r)]^2-m^2+m'^2 +\delta^2}\nonumber\\
&&\hspace*{3.8mm}=k^{*\mu} u_{\mu}(r),
\end{eqnarray}
\begin{equation}\label{BZ}
n'_{a,\mk'}=\frac{1}{\exp[(\Omega_{\mk'}(r)+\delta)/T(r)]-1},
\end{equation}
\begin{equation}\label{BZ1}
n'_{b,\mk'}=\frac{1}{\exp[(\Omega_{\mk'}(r)-\delta)/T(r)]-1},
\end{equation}
where $g_i$ is the degeneracy factor for hadron species $i$. $d^4\sigma_{\mu}(r)$ is the four-dimension element of freeze-out hypersurface,
and $u_{\mu}(r)$, $T(r)$ are the source four-velocity and the freeze-out temperature, respectively.
$k^{\mu}=(\omega_{\mk},{\mk})$ is the four-momentum of the particle, and ${\mk}'$
is the local-frame momentum corresponding to $\mk$. If the energy split between the boson and antiboson in the medium $\delta$ is 0, the $N_a(\mk)$
will be equal to $N_b(\mk)$.

The source distribution in our calculations is taken as
\begin{equation}
\label{rdis}
\rho(\mr)=C e^{-\mr^2/(2R^2)}\, \theta(r-2R),
\end{equation}
where $C$ and $R$ are the normalization constant and the source radius, respectively.
The expanding velocity of the source
is taken as
\begin{equation}
\mv(x)= \frac{u}{2R}\mr,
\end{equation}
where $u$ is a parameter. For the considered source and with the sudden freeze-out assumption, the single particle momentum distributions of boson and antiboson
become \cite{AsaCsoGyu99,Padula06,YZHANG_CPC15}
\begin{eqnarray}\label{SPs}
 N_a(\mk)\!&&=\!\int d^3 r\frac{g_i}{(2\pi)^3}\omega_{\mk}\rho(\mr)\, \! \Bigl\{|c'_{\mk'}|^2\,
n'_{a,\mk'}+\,|s'_{-\mk'}|^2\,[\,n'_{b,-\mk'}+1]\Bigr\},
\end{eqnarray}
\begin{eqnarray}\label{SPs1}
 N_b(\mk)\!&&=\!\int d^3 r\frac{g_i}{(2\pi)^3}\omega_{\mk}\rho(\mr)\, \! \Bigl\{|c'_{\mk'}|^2\,
n'_{b,\mk'}+\,|s'_{-\mk'}|^2\,[\,n'_{a,-\mk'}+1]\Bigr\}.
\end{eqnarray}

\section{Results}

\begin{figure}[htbp]
\includegraphics[scale=0.54]{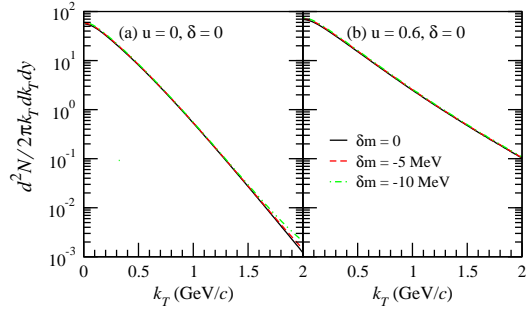}
\vspace*{-1mm}
\caption{(Color online) The transverse momentum spectra for Kaon with $\delta m$ = 0, -5 MeV and -10 MeV.}
\label{km}
\end{figure}

The transverse momentum spectra for Kaon with various mass-shift $\delta m$ ($\delta m = m'-m$) are shown in
Fig. \ref{km} (a) and (b). Here, the freeze-out temperature {\color{black}$T_f$} of Kaon is taken as 150 MeV \cite{Zhang_cpc19} and
the energy split between the boson and antiboson in the medium $\delta$ is taken as 0.
The mass-shift of Kaon in a hadronic medium is about -10 MeV at 150 MeV \cite{MFFK-PRL04}. In the calculation, the mass-shift of
Kaon is taken as -5 MeV and -10 MeV. {\color{black}The source radius $R$ is taken as 5 fm in this paper.}
The mass-shift leads to an increase in the yield of Kaon in large transverse momentum
regions. This effect increases with the increasing mass-shift and decreases with the increasing expanding velocity.
This phenomenon is similar to the effect of mass-shift on the transverse momentum spectra of $\phi$ meson \cite{Zhang20}.

\begin{figure}[htbp]
\includegraphics[scale=0.54]{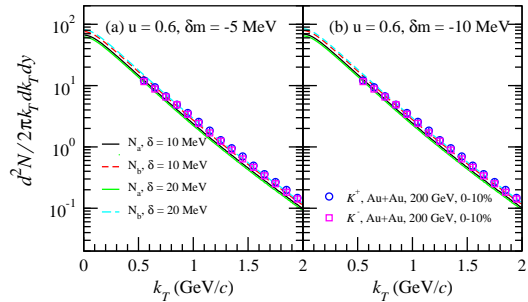}
\vspace*{-1mm}
\caption{(Color online) The transverse momentum spectra for Kaon and anti-Kaon with $\delta$ = 10 and 20 MeV {\color{black}for $T_f$ = 150 MeV.} {\color{black}The experimental data of the central Au + Au collision at $\sqrt{s_{NN}} =$ 200 GeV are measured by the PHENIX Collaboration \cite{PHENIX_Kaon}.}}
\label{kdet}
\end{figure}

In Fig. \ref{kdet} (a) and (b), we show the transverse momentum spectra for Kaon and anti-Kaon with $\delta$ = 10 and 20 MeV.
{\color{black}The experimental data of
the central Au + Au collision are measured by the PHENIX Collaboration \cite{PHENIX_Kaon}. The velocity parameter $u$ is taken
as 0.6 and it may be suitable for the central Au+Au collision at $\sqrt{s_{NN}} =$ 200 GeV.}
It can be seen that $\delta$ leads to a difference between the transverse momentum spectra of Kaon and anti-Kaon.
To analyze the difference quantitatively, we define a new quantity as
\begin{equation}
F\,=\,\frac{|N_a(\mk)-N_b(\mk)|}{N_a(\mk)+N_b(\mk)}.
\end{equation}

\begin{figure}[htbp]
\includegraphics[scale=0.54]{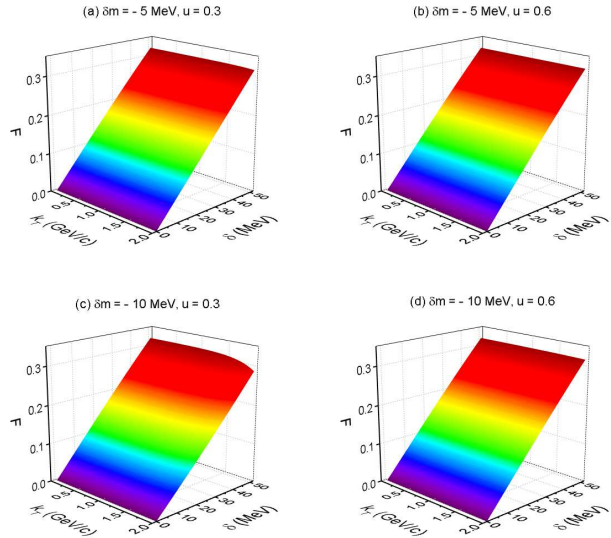}
\vspace*{-1mm}
\caption{(Color online) The quantity $F$ in $k_T-\delta$ plane for Kaon, {\color{black}where $T_f$ is taken as 150 MeV.}}
\label{k3d}
\end{figure}
\begin{figure}[htbp]
\includegraphics[scale=0.54]{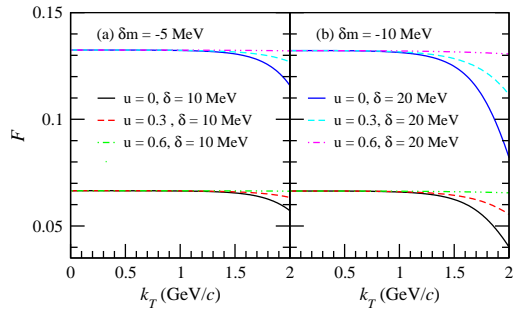}
\vspace*{-1mm}
\caption{(Color online) The quantity $F$ for Kaon with different expanding velocity, {\color{black}where $T_f$ is taken as 150 MeV.}}
\label{kptu}
\end{figure}

In Fig. \ref{k3d}, the quantity $F$ is shown in $k_T-\delta$ plane for Kaon. The difference between the transverse
momentum spectra of Kaon and anti-Kaon increases with the increasing $\delta$. The quantity $F$ for Kaon with different
expanding velocity is shown in Fig. \ref{kptu}. For fixed $\delta m$ and $\delta$, the difference between the transverse
momentum spectra of Kaon and anti-Kaon is almost the same in small momentum region ($k_T <$ 1.5 GeV) but decreases with
the increasing momentum in large momentum region($k_T >$ 1.5 GeV). In small momentum region ($k_T <$ 1.5 GeV), the difference
is not dependent on the expanding velocity. The difference increases with the increasing expanding velocity in large momentum
region($k_T >$ 1.5 GeV).

\begin{figure}[htbp]
\includegraphics[scale=0.54]{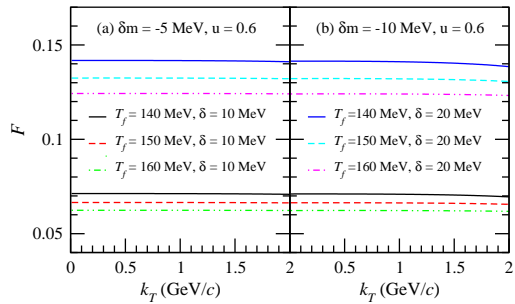}
\vspace*{-1mm}
\caption{(Color online) The quantity $F$ for Kaon with different freeze-out temperature.}
\label{kptf}
\end{figure}
{\color{black}Hadron mass shifts are caused by interactions in a dense medium and therefore vanish
on the freeze-out surface \cite{AsaCsoGyu99}. The interactions may increase with the increasing temperature of the medium and the
higher freeze-out temperature may lead to a greater mass difference (or energy difference) between boson and antiboson.}

{\color{black}The quantity $F$ for Kaon with different freeze-out temperature $T_f$ is shown in Fig. \ref{kptf}. For fixed $\delta$, the
difference between the transverse momentum spectra of Kaon and anti-Kaon decrease slightly with the increasing freeze-out temperature.
The energy difference between Kaon and anti-Kaon may increase with the increasing temperature of the medium. The
difference between the transverse momentum spectra of Kaon and anti-Kaon increase obviously with the increasing energy difference $\delta$.
Thus, the difference between the transverse momentum spectra of Kaon and anti-Kaon may increase with the increasing freeze-out temperature.}

\begin{figure}[htbp]
\includegraphics[scale=0.54]{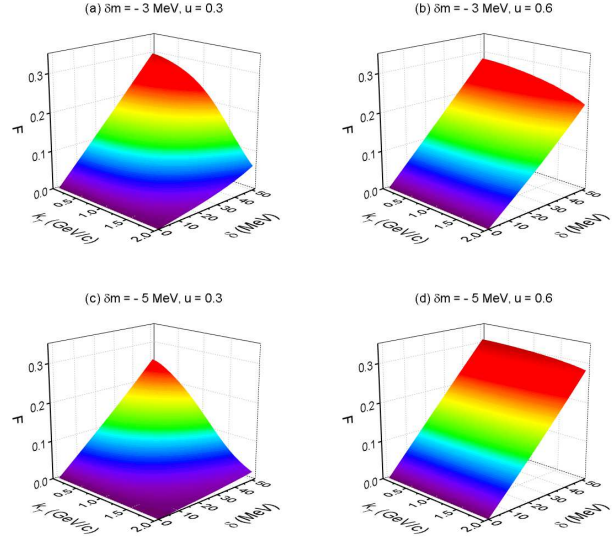}
\vspace*{-1mm}
\caption{(Color online) The quantity $F$ in $k_T-\delta$ plane for $D$ meson.}
\label{d3d}
\end{figure}

\begin{figure}[htbp]
\includegraphics[scale=0.58]{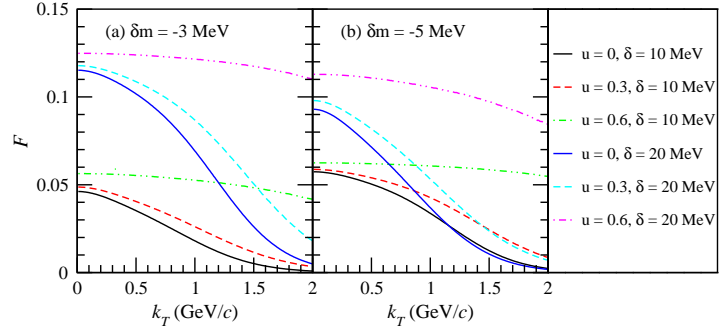}
\vspace*{-1mm}
\caption{(Color online) The quantity $F$ for $D$ meson with different expanding velocity {\color{black}for $T_f$ = 150 MeV.}}
\label{dptu}
\end{figure}

\begin{figure}[htbp]
\includegraphics[scale=0.54]{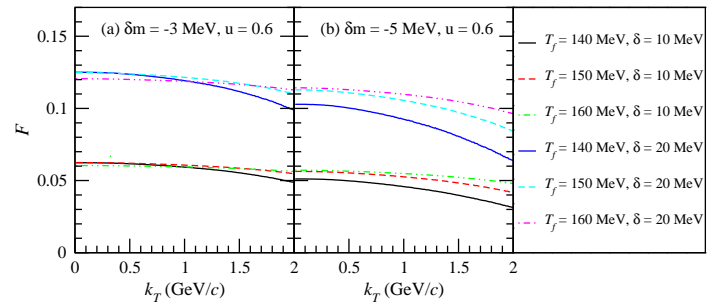}
\vspace*{-1mm}
\caption{(Color online) The quantity $F$ for $D$ meson with different freeze-out temperature.}
\label{dptf}
\end{figure}

In Fig. \ref{d3d}, the quantity $F$ is shown in $k_T-\delta$ plane for $D$ meson. The freeze-out temperature
of $D$ meson is taken as 150 MeV \cite{XPZ-19} and the $\delta m$ is taken as -3 MeV and -5 MeV \cite{XPZ-19,FMFK-PRC06}.
The difference between the transverse
momentum spectra of $D^+$ and $D^-$ increases with the increasing $\delta$ and decreases with the increasing momentum.
The quantity $F$ for $D$ meson with different expanding velocity is shown in Fig. \ref{dptu}.\,The difference increases
with the increasing expanding velocity.

{\color{black}The quantity $F$ for $D$ meson with different freeze-out temperature $T_f$ is shown in Fig. \ref{dptf}.
The difference between the transverse momentum spectra of $D^+$ and $D^-$ for $\delta m$ = -3 MeV is little affected by the
freeze-out temperature. For $\delta m$ = -5 MeV, the difference between the transverse momentum spectra of $D^+$ and $D^-$
increase slightly with the increasing freeze-out temperature for fixed $\delta$. Thus, with the increasing freeze-out temperature, the
energy difference increases and the difference between the transverse momentum spectra of $D^+$ and $D^-$ increases.}

Comparing the above results of $D$ meson and Kaon, the difference between the transverse
momentum spectra of boson and antiboson increases with the increasing expanding velocity of the source and decreases
with the increasing transverse momentum in large transverse mass region ($m_T\,=\,\sqrt{k_T^2\,+\,m^2}>\,1.6$ GeV).
{\color{black}The difference between the transverse momentum spectra of boson and antiboson may increase with the increasing freeze-out temperature.}

\section{Summary and discussion}
In high-energy heavy-ion collisions, the interactions of charged particles and
their corresponding antiparticles in a medium are different \cite{{LiLeeBrown-NPA97,SibTsuTho-EPJA99,CMKo-JPG01,
Mishra-PRC69-04,Mishra-PRC70-04,BlaCosKal-PRD12,Mishra-EPJA2019}}. Thus, the in-medium masses of charged particles and
their corresponding antiparticles are different.
In this paper, we study the effect of in-medium mass difference between a boson and an antiboson on their spectra.
The in-medium mass difference between a boson and an antiboson leads to a difference between the transverse
momentum spectra of boson and antiboson. This effect increases with the increasing in-medium mass difference between boson and antiboson.
The difference between the transverse momentum spectra of boson and antiboson
increases with the increasing expanding velocity of the source and decreases
with the increasing transverse momentum in large transverse mass region ($m_T\,>\,1.6$ GeV).
{\color{black}The interactions between the hadron and the medium may increase with the increasing temperature of the medium and the
higher freeze-out temperature may lead to a larger mass difference between boson and antiboson, and may give rise to a larger difference
between the transverse momentum spectra of boson and antiboson for higher freeze-out temperature.}

The transverse momentum spectra of particles are important observables in high-energy heavy-ion collisions.
The net-charge multiplicity distributions are applied to probe the signatures of the QCD phase transition and critical
point \cite{LAdamczyk,Xluo}. Based from above results, it is necessary to take the effect of in-medium mass difference between
a boson and an antiboson on their spectra in the calculations.

\begin{acknowledgments}
This research was supported by the National Natural Science Foundation
of China under Grant No. 11905085.
\end{acknowledgments}


\begin{thebibliography}{99}

\bibitem{STAR_PRL04_TMS}
J. Adams $et$ $al$. (STAR Collaboration), Phys. Rev. Lett. {\bf 92}, 112301 (2004).

\bibitem{PHENIX_PRC04_TMS}
S. S. Adler $et$ $al$. (PHENIX Collaboration), Phys. Rev. C {\bf 69}, 034909 (2004).

\bibitem{ALICE_PRL12_TMS}
B. Abelev $et$ $al$. (ALICE Collaboration), Phys. Rev. Lett. {\bf 109}, 252301 (2012).

\bibitem{ALICE_PRC13_TMS}
B. Abelev $et$ $al$. (ALICE Collaboration), Phys. Rev. C {\bf 88}, 044910 (2013).

\bibitem{AsaCso96}
M. Asakawa and T. Cs\"org\H o, Heavy Ion Physics {\bf 4}, 233 (1996);
hep-ph/9612331.

\bibitem{AsaCsoGyu99}
M. Asakawa, T. Cs\"org\H o and M. Gyulassy, Phys. Rev. Lett. {\bf 83},
4013 (1999).

\bibitem{Padula06}
S. S. Padula, G. Krein, T. Cs\"org\H{o}, Y. Hama, P. K. Panda, Phys.
Rev. C {\bf 73}, 044906 (2006).

\bibitem{Zhang15a}
Y. Zhang, J. Yang, W. N. Zhang, Phys. Rev. C {\bf 92}, 024906 (2015).

\bibitem{Zhang20}
Y. Zhang, J. Yang, Weihua Wu, Int. J. Mod. Phys. E {\bf 29}, 2050047 (2020).

\bibitem{LiLeeBrown-NPA97}
G. Q. Li, C. H. Lee, G. E. Brown, Nucl. Phys. A {\bf 625}, 372 (1997).

\bibitem{SibTsuTho-EPJA99}
A. Sibirtsev, K. Tsushima, and A. W. Thomas, Eur. Phys. J. A {\bf 6}, 351
(1999).

\bibitem{CMKo-JPG01}
C. M. Ko, J. Phys. G {\bf 27}, 327 (2001).

\bibitem{Mishra-PRC69-04}
A. Mishra, E. L. Bratkovskaya, J. Schaffer-Bielich, S. Schramm, and H. St$\ddot{o}$cker,
Phys. Rev. C {\bf 69}, 015202 (2004).

\bibitem{Mishra-PRC70-04}
A. Mishra, E. L. Bratkovskaya, J. Schaffer-Bielich, S. Schramm, and H. St$\ddot{o}$cker,
Phys. Rev. C {\bf 70}, 044904 (2004).

\bibitem{BlaCosKal-PRD12}
D. Blaschke, P. Costa, and Yu. L. Kalinovsky, Phys. Rev. D {\bf 85}, 034005
(2012).

\bibitem{Mishra-EPJA2019}
A. Mishra, A. K. Singh, N. S. Rawat, P. Aman, Eur. Phys. J. A {\bf 55}, 107
(2019).

\bibitem{XPZ-19}
P. Z. Xu, W. N. Zhang, Y. Zhang, Phys. Rev. C {\bf 99}, 011902(R) (2019).

\bibitem{YZHANG_CPC15}
Y. Zhang, J. Yang, and W.N. Zhang, Chin. Phys. C {\bf 39}, 034103 (2015).

\bibitem{Cooper}
F. Cooper and G. Frye, Phys. Rev. D {\bf 10}, 186 (1974).

\bibitem{KolHei03}
P. F. Kolb, U. Heinz, arXiv:nucl-th/0305084.

\bibitem{Kol00}
P. F. Kolb, J. Sollfrank, and U. Heinz, Phys. Rev. C {\bf 62}, 054909 (2000).

\bibitem{Zhang_cpc19}
Y. Zhang, J. Yang, Weihua Wu, Chin. Phys. C {\bf 43}, 074105 (2019).

\bibitem{MFFK-PRL04}
B. V. Martemyanov, A. Faessler, C. Fuchs, and M. I. Krivoruchenko, Phys. Rev.
Lett. {\bf 93}, 052301 (2004).

\bibitem{PHENIX_Kaon}
A. Adare $et$ $al$. (PHENIX Collaboration), Phys. Rev. C {\bf 88}, 024906 (2013).

\bibitem{FMFK-PRC06}
C. Fuchs, B. V. Martemyanov, A. Faessler, and M. I. Krivoruchenko, Phys. Rev.
C {\bf 73}, 035204 (2006).

\bibitem{LAdamczyk}
L. Adamczyk $et$ $al$. (STAR Collaboration), Phys. Rev. Lett. {\bf 113}, 092301 (2014).

\bibitem{Xluo}
X. Luo, Nucl. Phys. A {\bf 956}, 75 (2016).

\end{thebibliography}
\end{document}